\documentstyle[psfrag,aps,preprint,epsfig]{revtex}
\newcommand{\rem}[1]{{\bf #1}}
\newcommand{\eqref}[1]{eq.(\ref{#1})}

\newcommand{\Figref}[1]{Fig. \ref{#1}}

\begin{document}

\psfrag{psi1}{\LARGE $\Psi_1$}
\psfrag{psi2}{\LARGE $\Psi_2$}
\psfrag{psi3}{\LARGE $\Psi_3$}
\psfrag{phi1}{\LARGE $\Phi_1$}
\psfrag{phi2}{\LARGE $\Phi_2$}
\psfrag{phi3}{\LARGE $\Phi_3$}
\psfrag{hu}{\LARGE $H_u$}
\psfrag{hd}{\LARGE $H_d$}
\psfrag{q2}{\LARGE $Q_2$}
\psfrag{u2}{\LARGE $U^C_2$}
\psfrag{e2}{\LARGE $E^C_2$}
\psfrag{d1}{\LARGE $D^C_1$}
\psfrag{d2}{\LARGE $D^C_2$}
\psfrag{d3}{\LARGE $D^C_3$}
\psfrag{l1}{\LARGE $L_1$}
\psfrag{l2}{\LARGE $L_2$}
\psfrag{l3}{\LARGE $L_3$}
\psfrag{muy}{\Large $\mu y$}

\tighten

\preprint{TU-635}
\title{Proton Decay, Fermion Masses and  Texture 
from Extra Dimensions in SUSY GUTs}
\author{Mitsuru Kakizaki\footnote{e-mail: kakizaki@tuhep.phys.tohoku.ac.jp}
 and Masahiro Yamaguchi\footnote{e-mail: yama@tuhep.phys.tohoku.ac.jp}}
\address{Department of Physics, Tohoku University,
Sendai 980-8578, Japan}
\date{\today}
\maketitle
\begin{abstract}
In supersymmetry, there are gauge invariant dimension 5 proton decay operators
which must be suppressed by a mass scale much larger than the Planck mass. 
It is natural to expect that this suppression should
be explained by a mechanism that explains the hierarchical structure of the
fermion mass matrices. We apply this argument to the case where 
wave functions of chiral multiplets are localized under a kink background
along an extra spatial dimension and the Yukawa couplings as well as the 
coefficients of the proton decay operators are determined by the overlap 
of the relevant wave functions. A configuration is found in the context
of SU(5) supersymmetric grand unified theory that yields  realistic values of
quark masses, mixing angles, $CP$ phase and charged lepton masses and 
sufficiently small genuine dimension 5 proton decay operators. Inclusion of
SU(5) breaking effects is essential in order to obtain non-vanishing
$CP$ phase as well as correct lepton masses. The resulting
mass matrix has a texture structure in which texture zeros are a consequence
of extremely small overlap of the wave functions.  Our approach requires 
explicit breaking of supersymmetry in the extra dimension, which can 
be realized in  (de)constructing extra dimension.
\end{abstract} 

\clearpage

\section{Introduction}
One of the puzzles of modern particle physics is the origin of the structure
of masses and mixing in quarks and leptons. There have been many approaches
to explain the Yukawa couplings mostly based on some flavor symmetry.

In the supersymmetric (SUSY) extension of the Standard Model, besides the 
renormalizable Yukawa interaction, one can write non-renormalizable dimension
5 operators in the superpotential, which violate baryon and lepton numbers.
After dressed by superparticles, these operators induce nucleon decay 
\cite{protondecay}.  The
present experimental limit on the nucleon lifetime gives very stringent 
bounds on these operators, whose coefficients must be much smaller than the
inverse of the Planck scale for weak-scale SUSY breaking.  It is then natural
to expect that the mechanism which explains the fermion mass structure
will also explain the smallness of the genuine dimension 5 proton decay. 

Some time ago, Arkani-Hamed and Schmaltz \cite{Arkani-Hamed:2000dc}
proposed a new approach to these issues based on a fermion
localization mechanism in extra spatial dimensions.
\footnote{A similar proposal but with exponential Higgs configuration was 
given by Ref. \cite{Dvali:2000ha}.}
It is known that
chiral fermions are localized in solitonic backgrounds \cite{Jackiw:1976fn}.  
In Ref. \cite{Arkani-Hamed:2000dc}, 
the quarks and leptons have Gaussian
wave functions along one extra dimension under a kink background, and
the overlap of the wave functions of the two quarks (leptons)
determines the relevant Yukawa coupling. The small Yukawa couplings
are then attributed to the small overlap of the wave functions. 
Realistic configurations were given in Ref. \cite{Mirabelli:2000ks}.
\footnote{Issues of the fermion localization and the fermion masses were
discussed in the warped extra dimension 
in Refs. \cite{Grossman:2000ra,Chang:2000nh,Huber:2001ie}.}
In the
same way, the proton decay will be suppressed if the overlap of the
wave functions of three quarks and one lepton is small.

The purpose of this paper is to investigate whether this idea indeed
works in the context of supersymmetric grand unified theories (GUTs).
Issues of the fermion masses in SUSY GUTs using the idea of the
fermion localization were studied in Ref. \cite{Kaplan:2000av}. In addition
to the quark and lepton chiral multiplets, the Higgs supermultiplets
are also assumed to have the Gaussian wave functions along the extra
dimension and realistic masses and mixing angles of quarks are obtained.
Here we extend their approach to include the proton decay operators in
the analysis. We will show that one has to split the wave functions of the 
quarks in the same SU(5) multiplet in order to obtain sufficient $CP$ phase.
The splitting is realized by the mechanism proposed 
in Ref. \cite{Kakizaki:2001en}
(see also \cite{Maru:2001ch}).  Note that the same mechanism can realize the 
triplet-doublet Higgs mass splitting and the suppression of the
proton decay mediated by the triplet higgsino exchange. 

We shall obtain a configuration of the chiral multiplets in the
minimal supersymmetric standard model (MSSM), which is consistent with
the SU(5) GUT and explains the quark masses, the mixing angles and the
$CP$ phase, the charged lepton masses, and the suppression of the
genuine dimension 5 proton decay.  The resulting mass matrices of the
quarks and leptons have a variant of Fritzsch-type texture. In our approach,
texture zeroes are attributed to very small overlap of the wave functions. 

The organization of the paper is as follows.  In section 2, we review
the mechanism to localize the wave functions along the fifth
dimension. In section 3, we explain the basic ideas of our approaches
in the simple case where the wave functions have SU(5) invariant form. 
It turns out that $CP$ violation is too small in this case. Then in
section 4, we split the wave functions to obtain sufficiently large $CP$
phase. Here we need to contrive to suppress dimension 5 proton decay
operators. A workable configuration of the wave functions is given
there. In section 5, we argue various energy scales of our model and
discuss some subtle issues in this approach. The final section is devoted 
to conclusions and discussion.

\section{a brief review of localization mechanism}
We begin with a brief review of the localization mechanism 
of fields in the fifth dimension, which is based on 
the formalism of \cite{Kaplan:2000av,Arkani-Hamed:2001pv}.
Throughout this paper, 
we consider $5$D SU($5$) SUSY GUT with the fifth dimension $y$ compactified.
5D Planck scale is denoted by $M_*$ and the compactification scale by $M_c$.

We shall assume that 5D super-Poincare invariance is broken in some way
and only $N=1$ supersymmetry (in 4D sense) is respected. This allows us to 
write Yukawa interactions in the bulk in the form of superpotential, as we
will see shortly.   An appropriate formalism will be provided by utilizing
the idea of (de)constructing extra dimensions \cite{Arkani-Hamed:2001ca} 
which does not necessarily have
5D $N=1$ SUSY in the presence of matter interactions 
\cite{Csaki:2001em,Cheng:2001an}. 

Let us thus consider the Lagrangian
\begin{equation}
 {\cal L} =\int dy 
  \left\{ 
      \int d^4 \theta 
         \left( \Phi(y)^{\dagger}\Phi(y) +\Phi^C(y)^{ \dagger}\Phi^C(y)
               \right)
   +  \int d^2 \theta ~\Phi^C(y) \left[ \partial_y +M(y) \right] \Phi(y)
   + h.c. \right\}
 \label{eq:lagrangian}
\end{equation}
of $5$D spacetime.
Here, $\Phi$ is a $4$D $N=1$ SUSY chiral superfield and 
$\Phi^C$ is a charge conjugated chiral superfield, 
both of which combine to produce a $5$D $N=1$ SUSY hypermultiplet.  
$M(y)$ is decomposed as
\begin{equation}
       M(y) =\Xi(y)+M,
\end{equation}
where $\Xi(y)$ is the vacuum expectation value (vev) of 
a scalar component of a background chiral superfield
and $M$ is a mass parameter.  As we mentioned above, the interactions in the 
superpotential (\ref{eq:lagrangian}) break 5D $N=1$ SUSY (or $N=2$ in
4D SUSY).

Equations of motion for zero modes of 
scalar component $\phi$ and spinor one $\psi$ of $\Phi$ 
are obtained as
\begin{equation}
  \left( \partial_y +M(y) \right) \phi(y)=0, \ \
  \left( \partial_y +M(y) \right) \psi(y)=0,
\end{equation}
and for ones of $\Phi^C(y)$ as 
\begin{equation}
  \left( \partial_y -M(y) \right) \phi^C(y)=0, \ \
  \left( \partial_y -M(y) \right) \psi^C(y)=0.
\end{equation}
We assume that $\Xi$ has a kink configuration along the fifth dimension and 
approximate it as 
\begin{equation}
     \Xi(y)= 2 \mu^2 y, \quad \mu^2 > 0
     \label{eq:Xi}
\end{equation}
near the origin.
Thus, by appropriate boundary conditions 
we obtain the zero mode wave functions 
with Gaussian profile as
\begin{equation}
  \phi(y)=\psi(y)=
  \left(\frac{2 \mu^2}{\pi} \right)^{1/4}
  \exp\left[-\mu^2 (y-l)^2\right],
  \label{eq:gaussian}
\end{equation}
localized around $l\equiv -M/2 \mu^2$. 

If the extra dimension were non-compact, an anti-chiral zero mode would not
exist under the kink background because its wave function would not be 
normalizable. However in the compactified extra dimension, the wave function
will generically become normalizable, and one has to elaborate to forbid the
anti-chiral zero mode. Here zero modes for the anti-chiral fields, $\phi(y)^C$
 and
$\psi(y)^C$, are assumed to be absent. A possibility is that the anti-chiral
zero modes will be projected out by an orbifold boundary condition. 
Another possibility  to realize such a situation is to consider a 
$y$ dependent $Z$ (wave-function) factor in the 
Lagrangian (\ref{eq:lagrangian}), where the $Z$ factor vanishes at the
boundaries of the extra dimension. This makes the anti-chiral zero mode 
having non-normalizable wave function, and thus it does not survive. A detail 
on this realization will be given in Appendix A.
 
There are
also massive modes called Kaluza-Klein (KK) modes on top of these zero modes 
(massless modes). For the time being, we will focus on the zero modes
with Gaussian profile (\ref{eq:gaussian}).

\section{SU($5$) invariant configuration}
In this and the next sections we try to produce the hierarchy of the
fermion masses and very small proton decay operator
coefficients, using the localization mechanism described above.  The
important point is that smallness of these constants is a consequence
of small overlaps of wave functions.  The main purpose in the two
sections is to illustrate our procedure to determine localization
points of fields to reproduce fermion mass matrices and survive proton
decay constraints. We will obtain an almost realistic model except
that it has no $CP$ violating phase in the mass matrix. This model
nicely illustrates our procedure because the structure is rather simple and
also it suggests a hint how we should extend it to obtain 
non-vanishing $CP$ violation. 
We will give in
section 4 a more realistic model with $CP$ violation.

First, we consider the case that the MSSM fields which belong to the same 
multiplet in SU($5$) have the same wave functions localized at the same point 
along the extra dimension.
We assign the MSSM matter fields to the SU($5$) multiplets as usual:
\begin{eqnarray}
  \Psi(\rem{10}) &=& \frac{1}{\sqrt{2}} \left(
    \begin{array}{cc}
      U^C & Q\\
      - Q & E^C
    \end{array}
  \right) \nonumber \\
  \Phi(\rem{5}^*) &=& (D^C, L),
\end{eqnarray}
where $Q$ is the left-handed quark doublet, 
$U^C$ and $D^C$ are 
the charge conjugated right-handed up- and down-type quarks,
$L$ is the left-handed lepton doublet, and
$E^C$ is the charge conjugated right-handed charged lepton.
In the minimal case, Higgs multiplets are 
\begin{eqnarray}
  H(\rem{5}) &=& (H_T, H_u) \nonumber \\
  \bar{H}(\rem{5}^*) &=& (H_{\bar{T}}, H_d),
\end{eqnarray}
where $H_u$ and $H_d$ are the MSSM Higgses and $H_T$ and $H_{\bar{T}}$ are 
their color triplet partners.
Thus, the $5$D Yukawa couplings which leads to fermion masses are
\begin{equation}
  {\cal L} = 
  \int dy~d^2 \theta 
  \left\{ \frac{1}{4} \frac{f_U^{ij}}{\sqrt{M_*}} \Psi_i \Psi_j H 
    + \sqrt{2} \frac{f_D^{ij}}{\sqrt{M_*}} 
    \Psi_i \Phi_j \bar{H} \right\},
\end{equation}
where $f_{U,D}^{ij}$ are dimensionless 
coefficients and $i,j$ are family indices.
In non-minimal cases, Yukawa couplings take other form.
Any way, we assume that after SU($5$) breaking these couplings lead to
\begin{equation}
  \int dy~d^2 \theta 
  \left\{ \frac{f_U^{ij}}{\sqrt{M_*}} Q_i U^C_j H_u 
    + \frac{f_D^{ij}}{\sqrt{M_*}} Q_i D^C_j H_d 
  + \frac{f_L^{ij}}{\sqrt{M_*}} E^C_i L_j H_d \right\} .
  \label{eq:yukawa}
\end{equation}
For the moment, we concentrate on the quark sectors
and discard the wrong relations between down-type quark masses and 
charged lepton ones. 

Substituting zero mode wave functions of the form (\ref{eq:gaussian}) into 
\eqref{eq:yukawa} and integrating over the extra dimension, 
we obtain $4$D Yukawa couplings.
Here we denote Gaussian widths of the   matters (quarks and leptons), 
the up-type Higgs and the down-type Higgs by $\mu, \mu_{H_u}$ and $\mu_{H_d}$,
respectively, 
and define their ratios as $r_u \equiv \mu_{H_u}/\mu,r_d \equiv \mu_{H_d}/\mu$.
We assume that all matters have the same width for simplicity. Relaxing
this will not drastically change our main conclusions. Without
loss of generality, 
we set the location of $H_u$ at the origin.
Resulting $4$D up-type Yukawa coupling constants are 
\begin{equation}
  y_U^{ij} = F_U^{ij} \sqrt{\frac{r_u}{2 + r_u^2}} 
  \frac{2^{3/4}}{\pi^{1/4}} \exp 
  \left\{ - \frac{\mu^2}{2 + r_u^2} [ (1 + r_u^2)(l_i^2 + l_j^2) -2 l_i l_j ]
  \right\}, \quad F_U^{ij} \equiv f_U^{ij} \sqrt{\frac{\mu}{M_*}}.
  \label{eq:u_yukawa}
\end{equation}
Exponentially small coupling constants 
are a result of small overlap of the wave functions, which 
can be generated by at most $O(10)$ distance among fields relative to
typical magnitude of width.

Similarly, 
down-type Yukawa coupling constants are written as
\begin{equation}
  y_D^{ij} = F_D^{ij} \sqrt{\frac{r_d}{2 + r_d^2}} 
  \frac{2^{3/4}}{\pi^{1/4}} \exp 
  \left\{ - \frac{\mu^2}{2 + r_d^2} 
    [ (1 + r_d^2)(\tilde{l}_i^2 + \tilde{k}_j^2) -2 \tilde{l}_i \tilde{k}_j ]
  \right\}, \quad F_D^{ij} \equiv f_D^{ij} \sqrt{\frac{\mu}{M_*}}.
  \label{eq:d_yukawa}
\end{equation}
where $\tilde{l}_i \equiv l_i - l_{H_d}$, 
$\tilde{k}_i \equiv k_i - l_{H_d}$, and $k_i$ represents the location
of $\Phi(5^*)_i$. 

To obtain physical masses and mixings, 
we must translate fields from flavor basis into mass basis 
through unitary matrices as
\begin{eqnarray}
    U_Q^{u{\rm T}} y_U U_U &=& \hat{y}_U \equiv diag(y_u, y_c, y_t)\nonumber \\
    U_Q^{d{\rm T}} y_D U_D &=& \hat{y}_D \equiv diag(y_d, y_s, y_b)\nonumber \\
    U_E^{\rm T} y_L U_L &=& \hat{y}_L \equiv diag(y_e, y_\mu, y_\tau),
    \label{eq:ydiag}
\end{eqnarray}
from which we obtain the CKM matrix as 
\begin{equation}
  V_{\rm KM} = U_Q^{u \dag} U_Q^d.
  \label{eq:ckm}
\end{equation}
 
\subsection{Up-type Masses}
Next, we determine the locations of the fields using experimental data.
$5$D Yukawa coupling constants $f^{i,j}$ run from $M_*$ to $M_c$,
and then are matched to $4$D Yukawa coupling constants $y^{ij}$, and 
run again to low energy scale. 
However we ignore the running between  $M_*$ and $M_c$,
and rescale Yukawa coupling constants by
using one-loop MSSM renormalization group equations (RGE)
from the usual GUT scale $M_{\rm GUT} = 2 \times 10^{16}$ GeV to the EW scale,
by identifying that $M_c= M_{\rm GUT}$.  This simplification is
justified when we consider uncertainties of $f_U^{ij}$, $f_D^{ij}$
in the 5D Lagrangian. 

We adopt as experimental bounds the running quark masses evaluated at the 
$Z$-boson mass scale \cite{Fusaoka:1998vc,Groom:2000in}
\begin{eqnarray}
  m_u &=& 2.33^{+0.42}_{-0.45} ~\mbox{MeV}, \quad 
  m_c = 677^{+56}_{-61} ~\mbox{MeV}, \quad 
  m_t = 175 \pm 6 ~\mbox{GeV} \nonumber \\
  m_d &=& 4.69^{+0.60}_{-0.66} ~\mbox{MeV}, \quad 
  m_s = 93.4^{+11.8}_{-13.0} ~\mbox{MeV}, \quad 
  m_b = 3.00 \pm 0.11 ~\mbox{GeV} \nonumber \\
  m_e &=& 0.487 ~\mbox{MeV}, \quad
  m_\mu = 102.7 ~\mbox{MeV}, \quad
  m_\tau = 1.747 ~\mbox{GeV}, 
  \label{eq:mass_exp}
\end{eqnarray}
the magnitude of the CKM mixing matrix \cite{Groom:2000in}
\begin{equation}
  \left(
    \begin{array}{ccc}
      0.9742 \sim 0.9757 & 0.219 \sim 0.226 & 0.002 \sim 0.005 \\
      0.219 \sim 0.225 & 0.9734 \sim 0.9749 & 0.037 \sim 0.043 \\
      0.004 \sim 0.014 & 0.035 \sim 0.043 & 0.9990 \sim 0.9993
    \end{array}
  \right),
\end{equation}
and the Jarlskog invariant \cite{Atwood:2001jr}
\begin{equation}
  J \equiv V_{us} V_{cb} V_{ub}^* V_{cs}^* 
  = ( 1.8 \sim 3.1 ) \times 10^{-5},
\end{equation}
which represents $CP$ violation.

First we seek the locations of $\Psi_i$ from the observed up-type 
quark masses.
Since diagonal parts of $y_U$ contain the same field,
we expect that  $y_U$ is nearly diagonal.
Therefore in this case we can approximate locations of $\Psi_i$ as
\begin{equation}
  |\mu l_i| \sim \sqrt{ \frac{2 + r_u^2}{2 r_u^2} \log 
    \frac{F_U^{ii} v \sin \beta R_u}{\sqrt{2} m_{Ui}}}, 
  \quad R_u \equiv \sqrt{\frac{r_u}{2+r_u^2}} \frac{2^{3/4}}{\pi^{1/4}}
\end{equation}
where $v \sim 246 {\rm GeV}$ is the vacuum expectation value (vev) of the 
standard model Higgs 
and $\beta$ parameterizes the ratio of the MSSM Higgses' vevs as 
$\tan \beta \equiv \langle H_u \rangle / \langle H_d \rangle$.
Since $\sin \beta \sim 1$ for $\tan \beta \gtrsim 2$, 
$|\mu l_i|$ are sensitive to  $r_u$. 

There are two types of $\Psi$s' configuration which can lead to
realistic up-type quark masses.
One is (i) $\mu l_3 \sim 0 < \mu l_2 < \mu l_1$ type as in 
\cite{Kaplan:2000av}, 
and the other is (ii) $\mu l_1 < \mu l_3 \sim 0 < \mu l_2$ type, up to sign.
Recall that the wave function of $H_u$ is localized at the origin, and thus
the order one top Yukawa coupling requires  $\mu l_3 \sim 0$.

For example, in the case that $F_U^{ij} = 1, \tan \beta = 20$ and $r_u = 0.5$,
we obtain for the case (i), $\mu l_1 = 7.30, \mu l_2 = 5.10, \mu l_3 = 0.00$,
\begin{equation}
  m_u = 0.74 ~\mbox{MeV}, \quad m_c = 321 ~\mbox{MeV}, \quad m_t = 104~\mbox{GeV}, 
\end{equation}
and for the case (ii), $\mu l_1 = -7.20, \mu l_2 = 5.20, \mu l_3 = 0.60$,
\begin{equation}
  m_u = 1.03 ~\mbox{MeV}, \quad m_c = 255 ~\mbox{MeV}, \quad m_t = 96~\mbox{GeV}.
\end{equation}
Here we have used $U_Q^u = U_U \sim 1$ and the running quark masses are
those at  energy scale $M_{\rm GUT}$.
After RGE running to lower energy, these values can reside
in the range given in \eqref{eq:mass_exp}.

\subsection{Constraints From Proton Decay}
Although both of the two cases, (i) and (ii),
succeed in obtaining up-type quark masses,
they must also explain the null result of proton decay experiments.
We impose $R$ parity 
to forbid dimension $4$ proton decay throughout this paper,
and concentrate on genuine dimension $5$ proton decay operators suppressed by 
$5$D Planck scale. 
Ordinary dimension $5$ operators which arise from triplet Higgs exchange 
can be adequately suppressed by the mechanism described in 
\cite{Kakizaki:2001en}. Namely using the splitting of the wave functions of
the doublet and the triplet,  we can realize the configuration that 
the triplet Higgs fields are located far away
from the quarks and leptons, suppressing the Yukawa couplings of the
triplet to them and thus the proton decay through the triplet Higgs exchange. 

What we are concerned with are the genuine dimension $5$ proton decay operators
which are induced from the following interactions:
\begin{equation}
  {\cal L}_5 = \int dy~d^2 \theta \frac{\sqrt{2}}{4} \frac{d^{ijkl}}{M_*^2}
  (\Psi_i \Psi_j) (\Psi_k \Phi_l),
\end{equation}
where the brackets are contracted to the fundamental 
and the anti-fundamental representations, 
and $d^{ijkl}$ are constants which satisfy $d^{ijkl} = d^{jikl}$.
Since the gauge symmetry as well as the $R$-parity conservation does not
forbid this type of interactions, we expect that they have unsuppressed
coefficients $d^{ijkl} \sim O(1)$. 
By integrating over the fifth dimension, the above operators lead to
the genuine dimension $5$ proton decay operators
\begin{equation}
  {\cal L}_5 =
  \int d^2 \theta \left( \frac{1}{2} \frac{C_{Lf}^{ijkl}}{M_*} 
  Q_i Q_j Q_k L_l
  + \frac{C_{Rf}^{ijkl}}{M_*} E^C_i U^C_j U^C_k D^C_l \right),
  \label{eq:p_decay}
\end{equation}
in terms of $4$D fields.
Here $C^{ijkl}$ are constants proportional to $d^{ijkl}$:
\begin{equation}
  C_{L,Rf}^{ijkl} = \frac{D^{ijkl}}{\sqrt{\pi}} \exp 
  \left[ - \frac{\mu^2}{4} (3 p_j^2 + 3 p_k^2 + 3 q_l^2 
    - 2 p_j p_k - 2 p_k q_l - 2 p_j q_l ) \right], 
  \quad D^{ijkl} \equiv d^{ijkl} \frac{\mu}{M_*},
  \label{eq:pdecaycoe}
\end{equation}
where $p_j$ and $q_l$ denote $\Psi_j$'s and $\Phi_l$'s 
relative locations from $\Psi_i$, namely
$p_j \equiv l_j - l_i, q_l \equiv k_l - l_i$.
Note that $Q_i Q_i Q_i L_j$ and $E^C_i U^C_j U^C_j D^C_k$
identically vanish because of the Bose statistics.
What we hope is that the coefficients $C^{ijkl}$ become sufficiently
small due to small overlaps of the wave functions, which we will
study from now on.

In the above equation $C_{L,Rf}^{ijkl}$ are written in flavor basis, 
which are related to ones in mass basis as
\begin{eqnarray}
  C^{ijkl}_{Lm} &=& 
  {(U_Q^{u {\rm T}})^i}_m {(U_Q^{u {\rm T}})^j}_n 
    {(U_Q^{u {\rm T}})^k}_p {(U_L^{\rm T})^l}_q
  C^{mnpq}_{Lf} \nonumber \\
  C^{ijkl}_{Rm} 
  &=& {(U_E^{{\rm T}})^i}_m {(U_U^{{\rm T}})^j}_n 
    {(U_U^{{\rm T}})^k}_p {(U_D^{\rm T})^l}_q
  C^{mnpq}_{Rf}.
\end{eqnarray}

The decay mode $p \rightarrow K^+ \bar{\nu}$ gives the most severe constraints
on these constants.
The partial life time for this mode is larger than 
$ 1.9 \times 10^{33} {\rm yr}$ 
\cite{SK}, which implies \cite{Murayama:1994tc} for the first term of 
\eqref{eq:p_decay} (LLLL operators)
\begin{eqnarray}
  \sum_k \sqrt{|C_{Lm}^{iijk}|^2} 
  & \lesssim & C_c^{iijk}   \left( \frac{m_{\rm SUSY}}{1 {\rm~TeV}} \right)
  \left( \frac{0.03~{\rm GeV}^3}{|\beta_p|} \right)
  \left( \frac{10}{A_L} \right) \nonumber \\
  C_c^{112k} &=& 3.6 \times 10^{-11},\quad C_c^{221k} = 1.3 \times 10^{-10}
  \nonumber \\
  C_c^{113k} &=& 9.7 \times 10^{-10},\quad C_c^{331k} = 8.5 \times 10^{-8}
\label{eq:bounds}
\end{eqnarray}

if cancellation among coefficients in the proton decay amplitudes 
does not occur.
These values are those given at the GUT scale.
Here, we have assumed a common mass $m_{\rm SUSY}$ to 
all superparticles in the MSSM for simplicity,
and $\beta_p$ represents the hadronic matrix element parameter, which is
evaluated as 
\begin{equation}
  |\beta_p| = 0.003 \sim 0.03 ~{\rm GeV}^3
\end{equation}
by various methods \cite{beta}.
Hereafter we take $m_{\rm SUSY} = 1 ~{\rm TeV}$, and
the largest value $\beta_p = 0.03 ~{\rm GeV^3}$.
$A_L$ represents one loop renormalization effect due to gauge interaction from
$M_{\rm{GUT}}$ to $1~\rm{GeV}$. 
Here we neglect effect due to Yukawa interaction 
since it does not change our conclusions.

Similarly, for the second term of \eqref{eq:p_decay} (RRRR operators)
we have the following constraints:
\begin{eqnarray}
  |C_{Rm}^{ijkl}| 
  & \lesssim & C_c^{ijkl}
    \left( \frac{\sin 2 \beta}{0.10} \right)
  \left( \frac{m_{\rm SUSY}}{1~{\rm TeV}} \right)
  \left( \frac{0.03~{\rm GeV}^3}{|\alpha_p|} \right)
  \left( \frac{6.5}{A_R} \right) \nonumber \\
  C_c^{3311} &=& 4.7 \times 10^{-9},\quad C_c^{3211} = 4.5 \times 10^{-8}
\end{eqnarray}
at the GUT scale, where $\alpha_p$ is also the hadronic parameter
which satisfies $|\alpha_p| = |\beta_p|$ and $A_R$ is renormalization effect. 
The appearance of $\tan \beta$ is 
due to the fact that the proton decay diagram is generated
by Higgsino exchange, which inevitably depends on the Yukawa
couplings.

First we decide the location of $\Psi_i$ by considering constraints on 
the LLLL operators.
If $U_Q^u = 1$, we find
\begin{equation}
  \sum_l |C^{iijk}_{Lm}|^2 = \sum_l |C^{iijk}_{Lf}|^2.
\end{equation}
When $\Psi_i$ and $\Psi_j$ are localized close to each other, 
the overlap of their wave functions becomes large. To suppress the
proton decay, the localization point of the other one
$\Phi_k$ must be far away from them. Numerically we find
\begin{equation}
  q_k < q_-,  \quad ~\mbox{or} \quad q_+ < q_k, 
\end{equation}
where 
\begin{equation}
  \mu q_{\pm} \equiv \frac{1}{3} \left\{ \mu p_j 
    \pm \sqrt{12 \log \frac{D^{iijk}}{C_c^{iijk}\sqrt{\pi}} -8 
      \mu^2 p_j^2} \right\}.
\end{equation}
On the other hand, if $\Psi_i$ is localized far from $\Psi_j$, or
more explicitly   
\begin{equation}
    \mu |l_j - l_i| = \mu |p_j| > 
    \sqrt{\frac{3}{2} \log \frac{D^{iijk}}{C_c^{iijk} \sqrt{\pi}}},
\end{equation}
then we do not have any constraints on the position of  $\Phi_k$, and thus 
arbitrary $q_k$, namely $k_k$, is allowed.
Using the bounds (\ref{eq:bounds}) and taking $|D^{iijk}|=1$, we find if 
\begin{equation}
  \mu |l_1 - l_2| \gtrsim 5.9, \quad
  \mu |l_1 - l_3| \gtrsim 5.5, 
\label{eq:alwaysOK}
\end{equation}
then the proton decay induced by the LLLL operators is always suppressed to an
experimentally allowed level as far as $U_Q^U = 1$.

In the case (i), $\mu l_1 = 7.30, \mu l_2 = 5.10, \mu l_3 = 0.00$ do
not satisfy Eq.~(\ref{eq:alwaysOK}).  Thus the locations of the  $\Phi_k$ are
very constrained. For instance, when $|D^{iijk}| = 1$, the region
$1.2 < k_k < 11.4$ is prohibited even for $|\beta_p| = 0.003~\mbox{GeV}^3$.  
Even if we take $|D^{iijk}| = 10^{-2}$,
the corresponding excluded region is $1.9 < k_k < 10.7$.  Because of such large
excluded regions, it is difficult to obtain realistic
down-type quark masses and CKM parameters within at most $O(10)$
relative distance. Thus we will not consider the case (i) any more.

On the other hand, in the case (ii), 
$\mu l_1 = -7.20, \mu l_2 = 5.20, \mu l_3 = 0.60$ (with $|D^{iijk}| = 1$) 
satisfy the conditions
(\ref{eq:alwaysOK}), and thus we do not have any constraints on the locations
of
$\Phi_k$.

Next we consider the RRRR operators. If $U_U = 1$,
\begin{eqnarray}
  C^{ijkl}_{Rm} 
  &=& {(U_E^{{\rm T}})^i}_m {(U_D^{\rm T})^l}_q
  C^{mjkq}_{Rf}.
\end{eqnarray}
Even for large $\tan \beta$ constraints on these are weaker
than ones from the LLLL operators, since matter sector has SU($5$) symmetry.
Therefore, we expect that the
case (ii) survives the proton decay constraint.

In the above consideration we have taken $U_Q^u=U_U = 1$.
Thus the suppression of the proton decay rates only relies on the fact that 
$\Psi_1$ lives far from the other fields.
However, if $U_Q^u$ or $U_U$ are different from the unit matrix,
very rapid proton decay may be induced through off-diagonal elements, 
in particular from ${(U_Q^U)_i}^1$ and ${(U_U)_i}^1$.
We will come to this point later on.

\subsection{Texture}
Based on the aforementioned observations, we now seek
locations of $\Phi_i$ which generate realistic down-type quark masses and CKM
parameters.

The proton decay constraint prefers the case (ii) where $\Psi_1$ lives
on the opposite side of $\Psi_2$. It makes the structure of the mass
matrix very interesting as we will see soon.  Let us suppose that the
$\Phi(5^*)$ are placed in the following order along the extra dimension:
\begin{equation}
  \Psi_1 - \Phi_2 - \Psi_3 - \Phi_3 - \Psi_2 - \Phi_1 
  \label{eq:configuration}
\end{equation}
with Higgses localized around the third generation.
This can lead to a variant of Fritzsch type texture 
\cite{Fritzsch:1978vd} as follows \cite{Mirabelli:2000ks}
\begin{equation}
  y_D \simeq \left(
  \begin{array}{ccc}
    0 & a & 0 \\
    a' & 0 & b \\
    0 & d' & d 
  \end{array}
  \right)
  , \quad
  a, a \ll b \ll d, d', \quad a, a', b, d, d' > 0.
  \label{eq:fritzsch}
\end{equation}
Approximate zeros arise in the matrix when the overlaps of the wave functions
between $\Phi_i$ and $\Psi_j$ are extremely suppressed.

$y_D$ is diagonalized as \eqref{eq:ydiag}.
Since we have $U_Q^u \simeq 1$, 
we can approximate $V_{\rm CKM} \simeq U_Q^{d}$.
Therefore,
\begin{eqnarray}
  y_d &\sim& \frac{aa'd}{bd'}, \quad
  y_s \sim \frac{bd'}{\sqrt{d^{\prime 2} + d^2}}, \quad
  y_b \sim \sqrt{d^{\prime 2} + d^2} \nonumber \\
  |V_{us}| &\sim& \frac{ad}{bd'}, \quad 
  |V_{ub}| \sim \frac{ad'}{d^{\prime 2} + d^2}, \quad
  |V_{cb}| \sim \frac{bd}{d^{\prime 2} + d^2}.
  \label{eq:approximation}
\end{eqnarray}
This leads to a prediction
\begin{eqnarray}
    \frac{m_s^2}{m_b^2} \sim  \left| \frac{V_{ub}V_{cb}}{V_{us}} \right|
    \label{eq:prediction}
\end{eqnarray}
at $M_{\rm GUT}$. Observed values satisfy this relation very well.

\subsection{Quark Masses and Mixings}
We are now at the position to describe the configuration of the wave
function locations which generate realistic fermion masses and
mixings.

\begin{figure}[ht]
  \begin{center}
    \makebox[0cm]{
      \scalebox{0.9}{
        \includegraphics{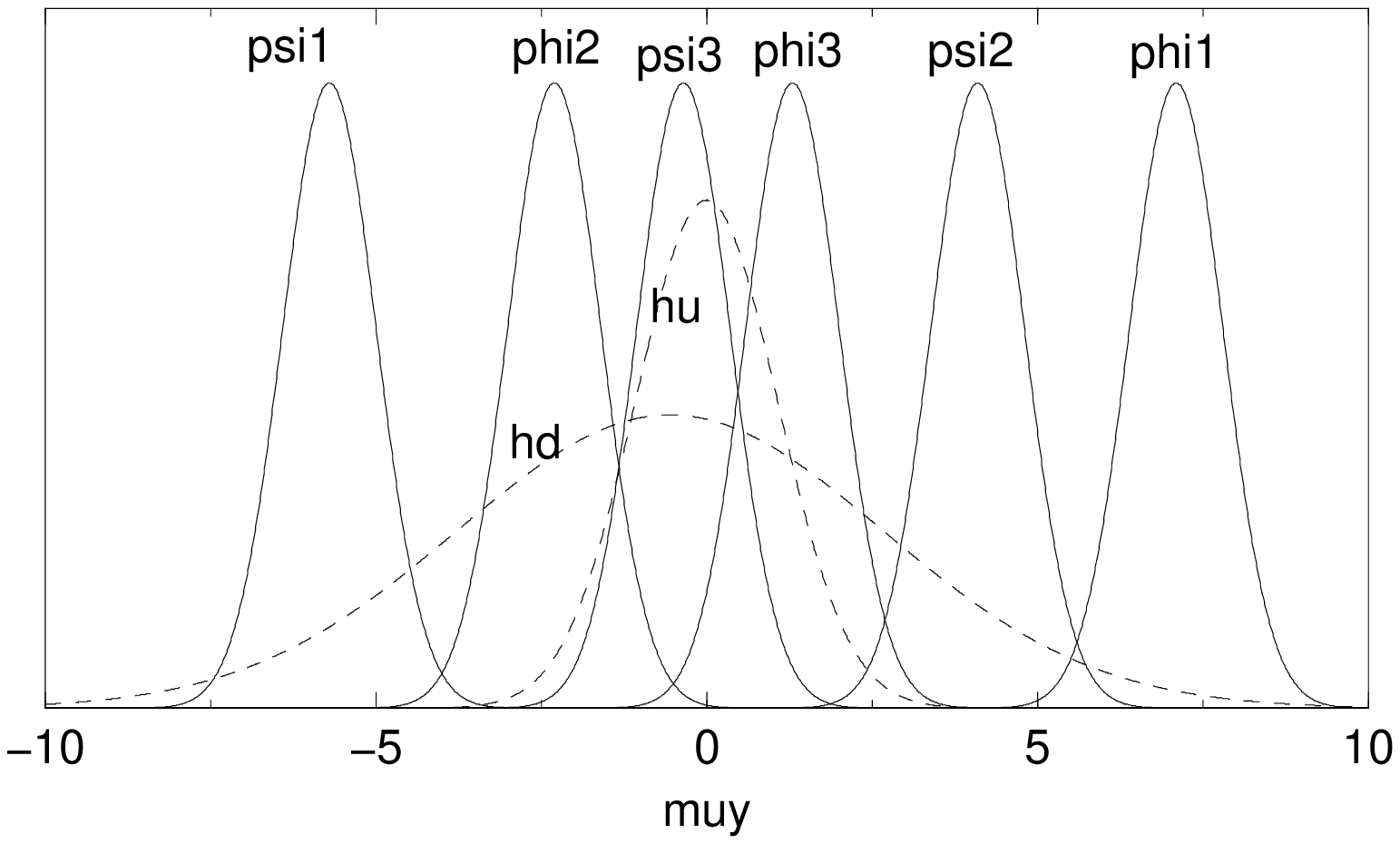}
        }
      }
    \caption{Profiles of fields which not only produce realistic quark masses 
      and mixing angles but also suppress genuine dimension $5$ proton decay.
      The resulting texture provides no $CP$ violation phase in the 
     CKM matrix.}
    \label{fig:profilegut}
  \end{center}
\end{figure}

For simplicity, we set $F_U^{ij} = F_D^{ij} = 1$ and $\tan \beta = 20$.
A choice  of the parameter set (see \Figref{fig:profilegut})
\begin{eqnarray}
  \quad r_u &=& 0.66, \quad r_d = 0.22, \quad \mu l_{H_d} = -0.55
  \nonumber \\
  \mu l_1 &=& -5.7, \quad \mu l_2 = 4.1, \quad \mu l_3 = -0.35
  \nonumber \\
  \mu k_1 &=& 7.1, \quad \mu k_2 = -2.3, \quad \mu k_3 = 1.3
\end{eqnarray}
yields Yukawa coupling matrices 
\begin{eqnarray}
  y_U = \left(
    \begin{array}{ccc}
      5.90 \times 10^{-6} &  7.31\times 10^{-22} & 1.52 \times 10^{-8} \\
      7.31 \times 10^{-22} & 1.61 \times 10^{-3} & 9.37 \times 10^{-6} \\
      1.52 \times 10^{-8} & 9.37 \times 10^{-6} & 0.629
    \end{array}
  \right) \nonumber \\
  y_D = \left(
    \begin{array}{ccc}
      1.02 \times 10^{-36} & 7.29 \times 10^{-4} & 8.33 \times 10^{-12} \\
      7.70 \times 10^{-4} & 4.78 \times 10^{-10} & 4.99 \times 10^{-3} \\
      1.77 \times 10^{-13} & 6.01 \times 10^{-2} & 0.101
    \end{array}
  \right). 
  \label{eq:y_gut}
\end{eqnarray}
These values give quark masses
\begin{eqnarray}
  m_u &=& 1.03 ~\mbox{MeV}, \quad m_c = 280 ~\mbox{MeV}, 
  \quad m_t = 110 ~\mbox{GeV} \nonumber \\
  m_d &=& 1.54 ~\mbox{MeV}, \quad m_s = 23.8 ~\mbox{MeV}, 
  \quad m_b = 1.02 ~\mbox{GeV}
\end{eqnarray}
and the magnitude of the mixing matrix
\begin{equation}
  \left(
    \begin{array}{ccc}
      0.975 & 0.221 & 0.003 \\
      0.220 & 0.975 & 0.036 \\
      0.011 & 0.035 & 0.999 \\
    \end{array}
  \right)
\end{equation}
at $M_{\rm GUT}$. 
After the RGE evolution, we obtain
\begin{eqnarray}
  m_u &=& 2.50 ~\mbox{MeV}, \quad m_c = 679 ~\mbox{MeV}, 
  \quad m_t = 176 ~\mbox{GeV} \nonumber \\
  m_d &=& 5.26 ~\mbox{MeV}, \quad m_s = 81.3 ~\mbox{MeV}, 
  \quad m_b = 2.98 ~\mbox{GeV}
\end{eqnarray}
and
\begin{equation}
  \left(
    \begin{array}{ccc}
      0.975 & 0.221 & 0.004 \\
      0.220 & 0.975 & 0.042 \\
      0.013 & 0.040 & 0.999 \\
    \end{array}
  \right)      
\end{equation}
at the $Z$-boson mass scale.
These are consistent with observed values. 

As for proton decay, including contribution from 
off-diagonal elements of unitary matrices 
we obtain
\begin{eqnarray}
  \sum_k \sqrt{|C_{Lm}^{112k}|^2} & & \sim 6.1 \times 10^{-15}, \quad
  \sum_k \sqrt{|C_{Lm}^{221k}|^2} \sim 6.8 \times 10^{-15} \nonumber \\
  \sum_k \sqrt{|C_{Lm}^{113k}|^2} & & \sim 4.1 \times 10^{-10}, \quad
  \sum_k \sqrt{|C_{Lm}^{331k}|^2} \sim 2.9 \times 10^{-9} \nonumber \\
  C_{Rm}^{3311} & & \sim 6.7 \times 10^{-10}, \quad
  C_{Rm}^{3211} \sim 1.0 \times 10^{-14},
\end{eqnarray}
which survives the current experimental bounds (\ref{eq:bounds}). Thus
this mechanism succeeds to explain not only quark masses and mixings
but also the suppression of the proton decay which arises from the
genuine dimension $5$ operators. To get small numbers, we used only at most
$O(10)$ parameters, {\it i.e.} some of $\mu l_i$ and $\mu k_i$ become
$\sim 10$. This is due to the Gaussian profiles of the wave functions
of the fields.

A similar argument can apply for $SO(10)$ GUT. However, it seems
difficult to explain proton decay and mass matrices simultaneously as
far as all matters in the same generation share a common wave
function.

One might worry that the configurations of \Figref{fig:profilegut} would
generate $\mu$ parameter (the supersymmetric mass of the
doublet Higgses) of order $M_*$.
This problem may be solved by 
introducing a singlet field $S$ localizing far from $SU(2)_L$ doublet Higgses
and considering the Yukawa interactions $S H_u H_d$ to yield a small
$\mu$ parameter 
\cite{Kakizaki:2001en}.

\subsection{Problems}
\label{subsec:cp}
Although we can obtain the realistic quark masses and mixings, the
above texture in which up-type Yukawa coupling matrix is diagonal 
cannot generate sufficient $CP$ violation \cite{Branco:2001rb} in the 
CKM matrix. 

Since up-type Yukawa coupling matrix is diagonal,
\begin{equation}
  y_D y_D^{\dag} = V_{KM}^* \hat{y}_D^2 V_{KM}^T
\end{equation}
On the other hand, \eqref{eq:fritzsch} implies
\begin{equation}
  y_D y_D^{\dag} = \left(
    \begin{array}{ccc}
      x & 0 & x \\
      0 & x & x \\
      x & x & x 
  \end{array}
  \right),
\end{equation}
where $x$s represent non-zero values.
Thus, we find
\begin{equation}
  y_d^2 V_{ud} V_{cd}^* + y_s^2 V_{us} V_{cs}^* + y_b^2 V_{ub} V_{cb}^* = 0.
\end{equation}
Using this relation and the unitarity of the CKM matrix, we obtain a relation
\begin{equation}
  \frac{y_s^2 - y_d^2}{y_b^2 - y_d^2} 
  = - \frac{V_{ub} V_{cb}^*}{V_{us} V_{cs}^*}.
\end{equation}
This is nothing but the prediction of \eqref{eq:prediction} and leads to
\begin{equation}
  J = Im(V_{us} V_{cb}V_{ub}^* V_{cs}^*) = 0,
\end{equation}
which means there is no $CP$ violation in the quark sector.
In fact, the Yukawa coupling matrices of 
\eqref{eq:y_gut} with arbitrary phases, 
give extremely small $CP$ violation, numerically $J \lesssim 1 \times 10^{-8}$.

Another problem is that this texture predicts equality of 
down-type masses and charged lepton masses at $M_{\rm GUT}$.
This is inevitable as far as GUT breaking Higgs does not couple with matters.

\section{SU($5$) breaking configuration}
In this section, we construct a model which explains the fermion
masses, mixings, $CP$ violation and the null results from proton decay
experiments at the same time. As we saw in the previous section, our
approach gave almost realistic mass matrices. The two difficulties
mentioned above are originated from the fact that the configurations
of the wave functions of the quarks and leptons have SU($5$) invariant
form, namely the fields belonging to the same multiplet have the same
wave function along the extra dimension.  Thus we will consider how
SU($5$) breaking is incorporated to improve these points.  Here, an
adjoint Higgs field $\Sigma$, which spontaneously breaks the SU($5$)
gauge symmetry, plays a crucial role in splitting the wave functions
of the quarks from ones of the leptons which belong to the same
multiplets \cite{Kakizaki:2001en}, so that we obtain the sufficient
$CP$ violation and the mass difference between down-type quarks and
charged leptons.

\subsection{Wave function Splitting Mechanism}
Let us here briefly review how the wave functions in a  SU($5$) multiplet
can be split after SU($5$) breaking. 
Besides the usual matters, $\Psi$ and $\Phi$, 
the model has  their conjugate fields, $\Psi^C$ and $\Phi^C$.
Thus we can write Yukawa interactions of matters 
with the adjoint Higgs $\Sigma$,
\begin{equation}
  {\cal L} = \int d^2 \theta \left\{ \Psi^C_i 
    \left[ \partial_y + M_i(y) - \frac{2}{3} g_i \Sigma \right] \Psi_i
    + \Phi^C_i 
    \left[ \partial_y + \bar{M}_i(y) 
      + \frac{1}{3} \bar{g}_i \Sigma \right] \Phi_i
    \right\}.
\end{equation}
Here, $i$ is a generation index, $g,\bar{g}$ are Yukawa coupling constants, 
and $M(y), \bar{M}(y)$ are position dependent $5$D masses including the 
$\Xi(y)$.
We assume that the adjoint Higgs $\Sigma$ develops the constant vev along the
extra dimension
\begin{equation}
  \langle \Sigma \rangle = V \left(
    \begin{array}{ccccc}
      2 &   &   &    &    \\
        & 2 &   &    &    \\
        &   & 2 &    &    \\
        &   &   & -3 &    \\
        &   &   &    & -3 \\
      \end{array}
  \right),
\end{equation}
by which the SU($5$) group is spontaneously broken. 
Therefore, after the GUT breaking, the above Lagrangian becomes of the
form
\begin{equation}
  {\cal L} = \int d^2 \theta F_i^C ( \partial_y + M_i + g_i V Y) F_i,
\end{equation}
where $F$'s are the MSSM fields, $Q, U^C, D^C, L$ and $E^C$, and 
$F^C$'s are their conjugated fields, and 
$Y$ represents the hypercharge of $F$ as
\begin{equation}
  \begin{array}{c|ccccc}
      F &   Q &  U^C & E^C & D^C & L  \\ 
      Y & 1/3 & -4/3 &   2 & 2/3 & -1,
  \end{array}
\end{equation}
and 
$M$ and $g$ include $\bar{M}$ and $\bar{g}$.
In this setting, the locations of zero mode wave functions of the MSSM fields 
are split as 
\begin{equation}
  l_i = - \frac{M_i + g_i Y V}{2 \mu^2}, \quad
  k_i = - \frac{\bar{M}_i + \bar{g}_i V Y}{2 \mu^2},
\end{equation}
with
\begin{equation}
  l(E^C) = 2 l(Q) - l(U^C).
  \label{eq:l(E^C)}
\end{equation} 
Since we no longer have SU($5$) symmetry in the matter sector,
this mechanism can explain the difference between down-type masses and 
charged lepton masses without using Georgi-Jarlskog mechanism
\cite{Georgi:1979df}.
However, the unification of $m_b$ and $m_\tau$ is accidental.

\subsection{$CP$ Violation}
The difficulty in generating sufficient $CP$ violation 
in previous SU($5$) invariant case
comes from the fact that the up-type Yukawa matrix is diagonal.
Thus we will consider a slight
deviation from the diagonal matrix for $y_U$ as follows:
\begin{equation}
  y_U = \left(
  \begin{array}{ccc}
    y_u & 0 & 0 \\
    0 & y_c & x \\
    0 & 0 & y_t 
  \end{array}
  \right) , \quad 
  y_D = \left(
  \begin{array}{ccc}
    0  & a  & 0 \\
    a' & 0  & b'e^{i \alpha} \\
    0  & d' & d
  \end{array}
  \right),
  \label{eq:nondiag}
\end{equation}
where complex phase only appears through the phase $\alpha$.
Since the above matrices have several zero entries,
it is sufficient to consider only one phase.
We keep elements $(y_U)^{1i}$ and $(y_U)^{i1}$ zero 
in order to avoid rapid proton decay.
Otherwise, proton decay coefficients in mass basis $C_{L,Rm}^{ijkl}$ may
become too large by rotating unsuppressed coefficients e.g. $C_{Lf}^{223k}$ 
by ${(U_Q^u)_3}^1$.

The unitary matrices which contribute to the CKM matrix are decomposed 
in orthogonal matrices $O$ and a phase matrix 
$P = diag(1, e^{- i \alpha}, 1)$ as
\begin{equation}
  U_Q^u = O_Q^u, \quad U_Q^d = P O_Q^d.
\end{equation}
These give the CKM matrix,
\begin{equation}
  V_{KM} = O_Q^{uT} P O_Q^d.
\end{equation}
For $O_Q^u = 1$ like in the previous SU($5$) invariant case, obviously $J = 0$.
On the other hand, for $O_Q^u \ne 1$ as in \eqref{eq:nondiag}
$CP$ violation compatible with experimental results can arise.

In the texture of \eqref{eq:nondiag}, 
following a similar argument given in section \ref{subsec:cp},
we obtain
\begin{equation}
  J \sim \frac{x}{y_t} \frac{m_b^2 - m_d^2}{m_s^2 - m_d^2} |V_{ub}|^2
  |V_{tb}| |V_{cb}| \sin \alpha.
\end{equation}
In order to produce $J \sim 10^{-5}$, 
at least $x/y_t \gtrsim 10^{-2}$ is required.

\subsection{Realistic Fermion masses, Mixings and $CP$ Violation}
Here we numerically describe a parameter set of models 
consistent with experimental results.
In searching parameter space,
for simplicity, we take $|F_{U,D}^{ij}| = 1$ as
\begin{equation}
  F_U = \left(
  \begin{array}{ccc}
    1 & 1 & 1 \\
    1 & 1 & 1 \\
    1 & 1 & 1 
  \end{array}
  \right) , \quad 
  F_D = \left(
  \begin{array}{ccc}
    1 & 1 & 1 \\
    1 & 1 & e^{i \alpha} \\
    1 & 1 & 1
  \end{array}
  \right),
\end{equation}
and fix $\tan \beta = 20$.
In order to produce the texture \eqref{eq:nondiag} and not to destroy 
the preferred features in the SU($5$) invariant case,
we simply split $\Psi_2$ and keep $\Psi_{1,3}$ unbroken.  

\begin{figure}[ht]
  \begin{center}
    \makebox[0cm]{
      \scalebox{0.9}{
        \includegraphics{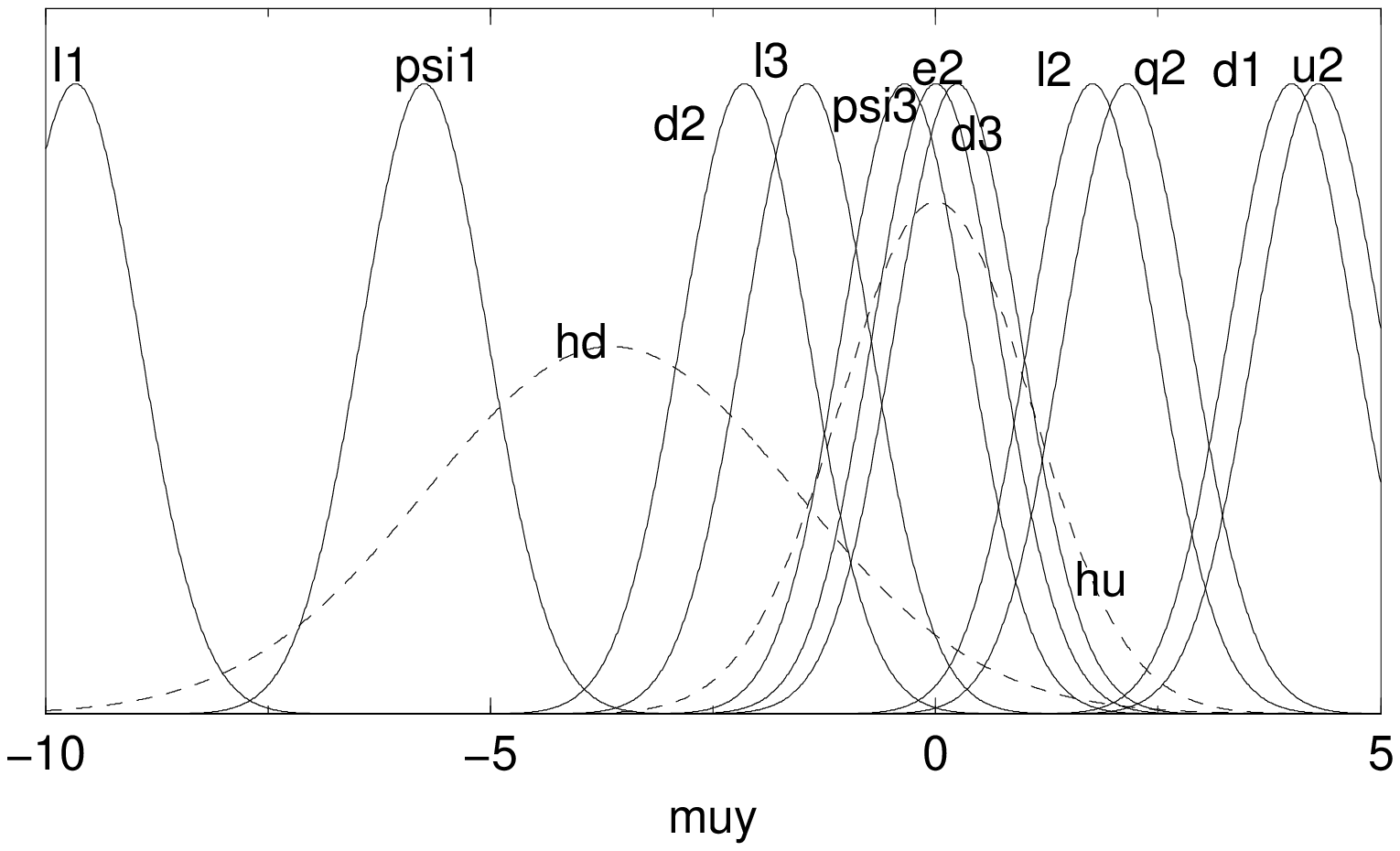}
        }
      }
    \caption{Profiles of fields which can produce realistic fermion masses, 
      CKM parameters and suppress genuine dimension $5$ proton decay. 
      Non-vanishing $CP$ phase in the CKM matrix is obtained in this case.}
    \label{fig:profilemssm}
  \end{center}
\end{figure}

An interesting parameter set we found by numerical survey is 
(see \Figref{fig:profilemssm}.)
\begin{eqnarray}
  r_u &=& 0.66, \quad r_d = 0.34, \quad \mu l(H_d) = -3.65 \nonumber \\
  \mu l(Q_1) &=& -5.75, \quad \mu l(Q_2) = 2.15, \quad \mu l(Q_3) = -0.35 
  \nonumber \\
  \mu l(U^C_1) &=& -5.75, \quad \mu l(U^C_2) = 4.30, \quad \mu l(U^C_3) = -0.35
  \nonumber \\
  \mu l(E^C_1) &=& -5.75, \quad \mu l(E^C_2) = 0.00, \quad \mu l(E^C_3) = -0.35
  \nonumber \\
  \mu k(D^C_1) &=& 4.00, \quad \mu k(D^C_2) = -2.15, \quad \mu k(D^C_3) = 0.24
  \nonumber \\
  \mu k(L_1) &=& -9.67, \quad \mu k(L_2) = 1.76, \quad \mu k(L_3) = -1.45,
\end{eqnarray}
which are derived from the following fundamental parameters,
\begin{eqnarray}
  g_1 V/\mu &=& 0, \quad g_2 V/\mu = 2.58, \quad g_3 V/\mu = 0, \nonumber \\ 
  \bar{g}_1 V/\mu &=& -16.404, \quad \bar{g}_2 V/\mu = 4.692, 
  \quad \bar{g}_3 V/\mu = -2.028, \nonumber \\ 
  M_1/\mu &=& 11.5, \quad M_2/\mu = -5.16, \quad M_3/\mu = 0.7, \nonumber \\ 
  \bar{M}_1/\mu &=& 2.936, \quad \bar{M}_2/\mu = 1.172, 
  \quad \bar{M}_3/\mu = 0.872
\end{eqnarray}
when $H_u$ is not shifted after GUT breakdown. From these values, 
we obtain the Yukawa coupling matrices as
\begin{eqnarray}
  y_U &=& \left(
  \begin{array}{ccc}
  4.81 \times 10^{-6} & 6.37 \times 10^{-23} & 1.10 \times 10^{-8} \\
  5.79\times 10^{-15} & 1.58 \times 10^{-3} & 2.16 \times 10^{-2} \\
  1.10 \times 10^{-8} & 3.29 \times 10^{-6} & 0.629
  \end{array}
  \right) , \nonumber \\
  y_D &=& \left(
  \begin{array}{ccc}
  4.97 \times 10^{-22} & 7.69 \times 10^{-4} & 7.50 \times 10^{-9} \\
  6.53 \times 10^{-4} & 1.14 \times 10^{-5} & 6.28 \times 10^{-3} e^{i \alpha} \\
  1.49 \times 10^{-6} & 5.34 \times 10^{-2} & 0.104
  \end{array}
  \right) , \nonumber \\
  y_L &=& \left(
  \begin{array}{ccc}
    3.84 \times 10^{-5} & 2.13 \times 10^{-13} & 4.89 \times 10^{-5} \\
    2.15 \times 10^{-21} & 1.14 \times 10^{-2} & 6.95 \times 10^{-2} e^{i \alpha} \\
    5.69 \times 10^{-20} & 6.88 \times 10^{-3} & 0.121
  \end{array}
  \right)
\end{eqnarray}
For $\alpha = 50^{\circ}$, the above parameter set predicts
\begin{eqnarray}
  m_u &=& 0.838 ~\mbox{MeV}, \quad m_c = 275 ~\mbox{MeV}, 
  \quad m_t = 110 ~\mbox{GeV} \nonumber \\
  m_d &=& 1.29 ~\mbox{MeV}, \quad m_s = 26.3 ~\mbox{MeV}, 
  \quad m_b = 1.02 ~\mbox{GeV} \nonumber \\
  m_e &=& 3.35 ~\mbox{MeV}, \quad m_\mu = 70.7 ~\mbox{MeV}, 
  \quad m_\tau = 1.22 ~\mbox{GeV},
\end{eqnarray} 
\begin{equation}
  \left(
    \begin{array}{ccc}
      0.975 & 0.222 & 0.003 \\
      0.221 & 0.974 & 0.037 \\
      0.010 & 0.035 & 0.999 
    \end{array}
  \right)
\end{equation}
and
\begin{equation}
  \quad J = 1.7 \times 10^{-5}
\end{equation}
at the $M_{\rm GUT}$ scale.
After the RGE evolution, 
\begin{eqnarray}
  m_u &=& 2.03 ~\mbox{MeV}, \quad m_c = 666 ~\mbox{MeV}, 
  \quad m_t = 176 ~\mbox{GeV} \nonumber \\
  m_d &=& 4.40 ~\mbox{MeV}, \quad m_s = 89.8 ~\mbox{MeV}, 
  \quad m_b = 2.96 ~\mbox{GeV} \nonumber \\
  m_e &=& 0.491 ~\mbox{MeV}, \quad m_\mu = 104 ~\mbox{MeV}, 
  \quad m_\tau = 1.75 ~\mbox{GeV},
\end{eqnarray}
\begin{equation}
  \left(
    \begin{array}{ccc}
      0.975 & 0.222 & 0.003 \\
      0.221 & 0.974 & 0.043 \\
      0.012 & 0.041 & 0.999 
    \end{array}
  \right)
\end{equation}
and 
\begin{equation}
  \quad J = 2.3 \times 10^{-5}
\end{equation}
at the $Z$-boson mass scale.
Furthermore, from \eqref{eq:pdecaycoe} we obtain
\begin{eqnarray}
  \sum_k \sqrt{|C_{Lm}^{112k}|^2} & & \sim 6.4 \times 10^{-13}, \quad
  \sum_k \sqrt{|C_{Lm}^{221k}|^2} \sim 1.3 \times 10^{-10} \nonumber \\
  \sum_k \sqrt{|C_{Lm}^{113k}|^2} & & \sim 1.9 \times 10^{-11}, \quad
  \sum_k \sqrt{|C_{Lm}^{331k}|^2} \sim 1.4 \times 10^{-9} \nonumber \\
  C_{Rm}^{3311} & & \sim 3.3 \times 10^{-10}, \quad
  C_{Rm}^{3211} \sim 3.0 \times 10^{-14}.
\end{eqnarray}
It turns out that the proton decay is suppressed to a level consistent with 
experimental limit.

As for the lepton sector, the locations of $E^C_i$ are fixed by 
\eqref{eq:l(E^C)}, and those of $L_i$ are adjusted to reproduce the observed 
charged lepton masses.

\section{scales of model}
Next we would like to clarify several points on the scales of this model.

So far, we have used the approximate form (\eqref{eq:Xi}) for
$\Xi(y)$,  which is valid  
near the origin. 
In this approximation, zero modes have Gaussian profiles (\eqref{eq:gaussian}),
and adding mass parameter $M$ to \eqref{eq:Xi} is merely
to shift wave functions along the fifth dimension.
However, this configuration will not hold in the entire extra dimension, but
rather it will behave for instance like
\begin{equation}
  \langle \Xi \rangle = \left\{
                  \begin{array}{ll}
                           \xi     & (y \geq L) \\
                           \mu^2 y & (-L < y < L) \\
                          -\xi     & (y \leq -L)
                   \end{array}
                      \right.
\end{equation}
with $\xi \equiv \mu^2 L$ for some $L$.  
A natural expectation is $\xi \sim M_*$. 
In order that all our previous arguments are justified, $L$ must be
sufficiently large: otherwise we would have different configurations for
wave functions. To obtain the fermion mass hierarchy we need
that $\mu L$ is at least order 10. Combined it with the expectation
$\xi \sim M_*$, we find
\begin{equation}
   \mu L \sim \frac{M_*}{\mu} \gtrsim 10.
\end{equation}

Another point we wish to clarify is behavior of KK modes.  
Wave functions for the KK modes obey the following eigenvalue equations
\begin{eqnarray}
 \left[ - \partial_y^2 + M^2(y) - \partial_y M(y) \right] L_n &=& m_n^2 L_n 
 \nonumber \\
 \left[ - \partial_y^2 + M^2(y) + \partial_y M(y) \right] R_n &=& m_n^2 R_n
  \label{eq:eq_of_motion}
\end{eqnarray}
where $L$ and $R$ represent components of chiral superfields and anti-chiral
super fields respectively, and $n$ labels the excitation level of a KK mode 
and $m_n$ is its mass .
For $m_n \ll \xi$, \eqref{eq:eq_of_motion} become
\begin{eqnarray}
  \left[ - \partial_y^2 + (2 \mu^2 y)^2 - 2 \mu^2 \right] L_n &=& m_n^2 L_n 
   \nonumber \\
  \left[ - \partial_y^2 + (2 \mu^2 y)^2 - 2 \mu^2 \right] R_n &=& m_n^2 R_n,  
\end{eqnarray}
and the wave functions are written in terms of Hermite polynomials.
Thus a KK mode with mass $m_n
\ll \xi$ is localized with narrow width.  A higher KK mode has a
more spread wave function, but is still  bounded.  On the contrary,
when $m_n > \xi$, since `energy' $m_n^2$ is always larger than
`potential' $M(y)^2 \pm \partial_y M(y)$, KK modes can freely
propagate all over the extra dimension.  Such KK modes, if exist in
the color triplet Higgses would be very dangerous, as they would
mediate unacceptably fast proton decay. To be safe, we should 
introduce cut-off below $M_*$ to eliminate such freely propagating
KK modes.\footnote{
Summation over the whole KK tower would, however, reproduce the 5D
picture in which the contribution to the proton decay from the heavy 
triplet is exponentially suppressed. We thank Martin Schmaltz for 
pointing it out. } 

The (de)constructing extra dimension naturally provides such a cut-off.
In this scenario, a (dynamical) scale $\Lambda$ is implemented in the 
theory. Below the energy scale $\Lambda$, the theory looks 5 dimensional with
one extra dimension. Above this scale, the extra dimension is resolved and
the theory looks 4 dimensional with gauge group $SU(5)^N$ with $N$ being some
large integer.  In this setup, it is very natural to expect that there is 
no KK mode above $\Lambda$, which provides the cut-off we want to have.

Furthermore in the (de)constructing extra dimension,  gravity does not
propagate this extra dimension, so that we can identify the 5D Planck scale
$M_*$ with the 4D (reduced) Planck scale $2.4 \times 10^{18} $ GeV. 

Summarizing these arguments, we find that a set of the parameters as
\begin{eqnarray}
      M_* &\sim & 10^{18} ~\mbox{GeV} \nonumber \\
      \mu &\sim & 10^{17} ~\mbox{GeV} \nonumber \\
      L^{-1} & \sim & 10^{16} ~\mbox{GeV}
\end{eqnarray}
provide an example of the setting in which our mechanism works.

Perturbativity of 5D gauge interactions requires 
\begin{equation}
            N_{KK} \frac{g^2}{16 \pi^2} \lesssim 1,
\label{eq:perturbation}
\end{equation}
where $g$ is a gauge coupling constant and $N_{KK}$ is the number of
the KK modes below the cut-off $\Lambda$. We expect that
\begin{equation}
  N_{KK} \sim \Lambda L \lesssim M_* L \sim 100,
\end{equation}
and thus with the gauge coupling of order unity, eq.~(\ref{eq:perturbation})
is satisfied. 

\section{Conclusions and discussion}
In this paper, we have pursued a very natural expectation that the 
mechanism which explains the masses and mixing of the quarks and leptons
should also explain the smallness of the dangerous genuine dimension 5
proton decay operators in supersymmetry. This philosophy provides a 
test of models of flavor and here we applied to the mechanism of localizing
fermions under kink background along the extra dimension.   
We showed that the localization mechanism can provide a successful 
configuration of the wave functions of the chiral multiplets which is 
consistent with the SUSY SU(5) and yields the realistic fermion mass 
structure and suppresses the dimension 5 proton decay.

Here we would like to summarize our ingredients  on the localization mechanism.
\begin{itemize}
\item The mechanism advocated by Arkani-Hamed and Schmaltz is used. In
this approach, there exists a singlet filed $\Xi$ which has a kink 
configuration along the extra dimension. Bulk fermions are assumed to 
have  Yukawa interactions with the singlet in the bulk, 
which is a necessary ingredient
of this localization mechanism. The width and the localization of the
fermion wave function are controlled by the Yukawa coupling and the invariant
mass of the fermion. In the context of  SUSY,  $D=5$ full SUSY would
prevent such Yukawa interactions in the superpotential. We argue that
in the deconstructing extra dimension $D=5$ SUSY can be explicitly broken
to $D=4$ $N=1$ SUSY, allowing the desired superpotential.

\item
In order to obtain a chiral theory, one needs to elaborate to kill anti-chiral
zero modes. One way would be to consider a $Z_2$ orbifold compactification, 
in which appropriate boundary conditions would allow only chiral zero modes.
An alternative way is to consider a non-trivial $Z$ factor for the fermion
kinetic terms, in which $Z$ vanishes at the boundaries. In this case, the
anti-chiral zero modes have non-normalizable wave functions and thus they
do not exist.
\end{itemize}

To obtain a realistic configuration, one further requires 
\begin{itemize}
\item Splitting of wave functions in a single SU(5) multiplet is required
to have a non-vanishing CKM phase.

\item Appropriate couplings and masses should  be chosen to reproduce the
fermion masses and mixing and to suppress the proton decay to an 
experimentally acceptable level. 
\end{itemize}

In this paper, we extensively discussed the SU(5) GUT. It is interesting
to consider the case of SO(10) and larger groups. Here we  briefly
mention the SO(10) case. As is well-known, all quarks and leptons as well
as a right-handed neutrino can be  embedded into a single 16 dimensional 
representation in SO(10). These fields in the same multiplet are localized
in different points after the SO(10) breaking.
To see this point closely, let us consider the following  symmetry
breaking chain 
$SO(10) \to SU(5) \to SU(3) \times SU(2) \times U(1)$. In the first symmetry
breaking, 10, $\bar 5$, and $1$ (right-handed neutrino) in SU(5) 
terminology are split. Thus our arguments given for the SU(5) case 
can apply to this case. A crucial difference is that 
the positions and widths of the right-handed neutrinos are not freely chosen, 
but are related to those of 10 and  $\bar 5$ because the splitting is 
attributed to the expectation value of the Higgs responsible for the breaking 
$SO(10) \to SU(5)$. It is interesting to 
see whether the realistic 
neutrino masses and mixing are realized in this case, which is however
beyond the scope of this paper.

To conclude, the idea of the fermion localization along the extra
dimension can pass the phenomenological test of models of flavor.  Our 
explicit construction shows that this idea can indeed work. Of course,
ours is just one possibility among divergent approaches to the extra 
dimensions. Further investigation along this line should be encouraged. 

\

 After completion of this work, we received a preprint \cite{Kaplan:2001ga} 
which deals with a similar subject.

\section*{Acknowledgment}               
This work was supported in part by the
Grant-in-aid from the Ministry of Education, Culture, Sports, Science
and Technology, Japan, priority area (\#707) ``Supersymmetry and
unified theory of elementary particles,'' and in part by the
Grants-in-aid No.11640246 and No.12047201.

\appendix

\section{A Mechanism to Obtain Chiral Zero Modes}
In this appendix, we describe a mechanism to 
obtain only the zero modes of a chiral superfield $\Phi$ while 
forbidding those of a charge conjugated chiral superfield $\Phi^C$
in five dimensions with the extra dimension compactified on the
orbifold $S_1/Z_2$.
The key point for this mechanism to work
is introduction of a $Z$ factor with a non-trivial profile 
along the extra dimension. Here, we assume
that the $Z$ factor is nearly constant in the bulk 
but vanishes at the boundaries of the 
extra dimension. Here we do not specify the origin of the
$Z$ factor.

Instead of  \eqref{eq:lagrangian}, we thus suppose the following 
five-dimensional action:
\begin{eqnarray}
  S & = & \int \! d^5x \left\{ \int \! d^4 \theta 
    Z(y) (\Phi^{\dag} \Phi + \Phi^{C\dag} \Phi^C) \right. \nonumber \\
  & & \left. + \int \! d^2 \theta \left( \frac{1}{2} Z(y) (\Phi^C \stackrel{\leftrightarrow}{\partial_y} \Phi) + M(y) \Phi^C \Phi  + h.c. \right) \right\},
\end{eqnarray}
where $y$-derivative does not act on $Z(y)$.

We expand  the fields into their KK modes,
\begin{eqnarray}
  \Phi = \frac{1}{\sqrt{Z}} \sum_{n=0}^\infty f_L^{(n)}(y) \Psi^{(n)}(x),
  & \quad &
  \Phi^C = \frac{1}{\sqrt{Z}} \sum_{n=0}^\infty f_R^{(n)*}(y) \Psi^{C(n)}(x),
\end{eqnarray}
where $\{f_L^{(n)}\}$ and $\{f_R^{(n)}\}$ span  complete 
 orthonormal bases as follows:
\begin{eqnarray}
  \int \! dy ~f_L^{(m)*} f_L^{(n)} = \delta_{mn}, \quad
  \int \! dy ~f_R^{(m)*} f_R^{(n)} = \delta_{mn}.
\end{eqnarray}

Integrating over the extra dimension, we obtain
\begin{eqnarray}
  S & = & \sum_{n=0}^{\infty} \int \! d^4x \left\{ \int \! d^4 \theta 
    \left( \Psi^{(n)\dag} \Psi^{(n)} + \Psi^{C(n)\dag} \Psi^{C(n)} \right)
    + \int \! d^2 \theta \left( m_n \Psi^{C(n)} \Psi^{(n)} 
      + h.c. \right) \right\},
\end{eqnarray}
if the wave functions $f_{L,R}^{(n)}$  satisfy the following eigenvalue
equations:
\begin{eqnarray}
  \left( \partial_y + \frac{M(y)}{Z(y)} \right)f_L^{(n)} = m_n f_R^{(n)}, \quad
  \left( - \partial_y + \frac{M(y)}{Z(y)} \right) f_R^{(n)} = m_n f_L^{(n)}.
\end{eqnarray}
Notice that the mass term $M(y)$ is practically replaced with $M(y)/Z(y)$. 
In particular, the equations for the zero modes are written
\begin{eqnarray}
  \left( \partial_y + \frac{M(y)}{Z(y)} \right)f_L^{(0)} = 0, \quad
  \left( - \partial_y + \frac{M(y)}{Z(y)} \right) f_R^{(0)} = 0,
\end{eqnarray}
whose general solutions are easily obtained as
\begin{eqnarray}
  f_L^{(0)} (y)
  = N_L \exp \left( - \int_0^y dy' \frac{M(y')}{Z(y')} \right), \quad 
  f_R^{(0)} (y)
  = N_R \exp \left( \int_0^y dy' \frac{M(y')}{Z(y')} \right),
\end{eqnarray}
where $N_{L,R}$ are normalization constants. 
Since $M(y)/Z(y)$ diverges at the boundaries of the extra dimension,
one of the zero-mode wave functions, say $f_R^{(0)}$, is not normalizable. 
Thus there exists only the chiral zero mode for $\Phi$. Notice that 
the shape of the zero mode  $f_L^{(0)}$ is  not changed drastically 
compared with the case $Z=1$.

%


\begin{references}


\bibitem{protondecay}
N.~Sakai and T.~Yanagida,
Nucl.\ Phys.\ B {\bf 197} (1982) 533.

S.~Weinberg,
Phys.\ Rev.\ D {\bf 26} (1982) 287.


\bibitem{Arkani-Hamed:2000dc}
N.~Arkani-Hamed and M.~Schmaltz,
Phys.\ Rev.\ D {\bf 61} (2000) 033005
[hep-ph/9903417].




\bibitem{Dvali:2000ha}
G.~R.~Dvali and M.~A.~Shifman,
Phys.\ Lett.\ B {\bf 475} (2000) 295
[hep-ph/0001072].


\bibitem{Jackiw:1976fn}
R.~Jackiw and C.~Rebbi,
Phys.\ Rev.\ D {\bf 13} (1976) 3398.


\bibitem{Mirabelli:2000ks}
E.~A.~Mirabelli and M.~Schmaltz,
Phys.\ Rev.\ D {\bf 61} (2000) 113011
[hep-ph/9912265].


\bibitem{Grossman:2000ra}
Y.~Grossman and M.~Neubert,
Phys.\ Lett.\ B {\bf 474} (2000) 361
[hep-ph/9912408].



\bibitem{Chang:2000nh}
S.~Chang, J.~Hisano, H.~Nakano, N.~Okada and M.~Yamaguchi,
Phys.\ Rev.\ D {\bf 62} (2000) 084025
[hep-ph/9912498].


\bibitem{Huber:2001ie}
S.~J.~Huber and Q.~Shafi,
Phys.\ Lett.\ B {\bf 498} (2001) 256
[hep-ph/0010195].


\bibitem{Kaplan:2000av}
D.~E.~Kaplan and T.~M.~Tait,
JHEP{\bf 0006} (2000) 020
[hep-ph/0004200].


\bibitem{Kakizaki:2001en}
M.~Kakizaki and M.~Yamaguchi,
[hep-ph/0104103].


\bibitem{Maru:2001ch}
N.~Maru,
[hep-ph/0108002].


\bibitem{Arkani-Hamed:2001pv}
N.~Arkani-Hamed, L.~Hall, D.~Smith and N.~Weiner,
Phys.\ Rev.\ D {\bf 63} (2001) 056003
[hep-ph/9911421].


\bibitem{Arkani-Hamed:2001ca}
N.~Arkani-Hamed, A.~G.~Cohen and H.~Georgi,
Phys.\ Rev.\ Lett.\  {\bf 86} (2001) 4757
[hep-th/0104005].


\bibitem{Csaki:2001em}
C.~Csaki, J.~Erlich, C.~Grojean and G.~D.~Kribs,
[hep-ph/0106044].


\bibitem{Cheng:2001an}
H.~C.~Cheng, D.~E.~Kaplan, M.~Schmaltz and W.~Skiba,
Phys.\ Lett.\ B {\bf 515} (2001) 395
[hep-ph/0106098].


\bibitem{Fusaoka:1998vc}
H.~Fusaoka and Y.~Koide, 
Phys.\ Rev.\ {\bf D 57}, 3986 (1998)
[hep-ph/9712201].


\bibitem{Groom:2000in}
D.~E.~Groom {\it et al.}  [Particle Data Group Collaboration],
Eur.\ Phys.\ J.\ C {\bf 15} (2000) 1.


\bibitem{Atwood:2001jr}
D.~Atwood and A.~Soni,
Phys.\ Lett.\ B {\bf 508} (2001) 17
[hep-ph/0103197].


\bibitem{SK}
Y.~Totsuka  [Super-Kamiokande collaboration], 
talk presented at SUSY2K, CERN, Geneva, Switzerland, 26 June -1 July 2000.


\bibitem{Murayama:1994tc}
H.~Murayama and D.~B.~Kaplan,
Phys.\ Lett.\ B {\bf 336} (1994) 221
[hep-ph/9406423].

\bibitem{beta}
B.~L.~Ioffe,
Nucl.\ Phys.\ B {\bf 188} (1981) 317
[Erratum-ibid.\ B {\bf 191} (1981) 591].

Y.~Tomozawa,
Phys.\ Rev.\ Lett.\  {\bf 46} (1981) 463
[Erratum-ibid.\  {\bf 49} (1981) 507].

J.~F.~Donoghue and E.~Golowich,
Phys.\ Rev.\ {\bf D 26} (1982) 3092.

S.~J.~Brodsky, J.~Ellis, J.~S.~Hagelin and C.~Sachrajda,
Nucl.\ Phys.\ {\bf B238} (1984) 561.

S.~Aoki {\it et al.}  [JLQCD Collaboration],
Phys.\ Rev.\ D {\bf 62} (2000) 014506
[hep-lat/9911026].


\bibitem{Fritzsch:1978vd}
H.~Fritzsch,
Phys.\ Lett.\ B {\bf 73} (1978) 317;


\bibitem{Branco:2001rb}
G.~C.~Branco, A.~de Gouvea and M.~N.~Rebelo,
Phys.\ Lett.\ B {\bf 506} (2001) 115
[hep-ph/0012289].


\bibitem{Georgi:1979df}
H.~Georgi and C.~Jarlskog,
Phys.\ Lett.\ B {\bf 86} (1979) 297.


\bibitem{Kaplan:2001ga}
D.~E.~Kaplan and T.~M.~Tait,
[hep-ph/0110126].

\end{references}
\end{document}